\documentstyle[aps,twocolumn,floats,epsf]{revtex} 
\begin{document}  
\draft
\title{Spectral Statistics: From Disordered to Chaotic Systems}                               
\author{Oded Agam}  
\address{Department of Physics, Technion, Haifa 32000, Israel}
\author{Boris L. Altshuler and Anton V. Andreev} 
\address{NECI, 4 Independence Way, Princeton, NJ\ 08540, USA,\\
Department of Physics, Massachusetts Institute of Technology, 77
Massachusetts Avenue, Cambridge, MA\ 02139, USA }
\maketitle     
\begin{abstract}  

The relation between disordered and chaotic systems is investigated.
It is obtained by identifying the diffusion operator of the disordered
systems with the Perron-Frobenius operator in the general case.
This association enables us to extend results obtained in the diffusive
regime to general chaotic systems. In particular, the two--point level 
density correlator and the structure factor  for general chaotic systems 
are calculated and characterized.  The behavior of the structure 
factor around the Heisenberg time is quantitatively described in terms 
of short periodic orbits.

\end{abstract}                                                                  
\par                                                                            
\vspace{0.5cm}  
The statistical description of the quantum spectra of systems
which exhibit chaotic dynamics in their classical limit, 
has been conducted mainly along two routes. One is to study
an {\em ensemble} of similar systems, such as disordered metallic 
grains, where electrons experience scattering by a random potential.
In this approach, ensemble averaging is a crucial step done at 
an early stage of the calculation. The results of such a 
calculation apply to an individual member of the ensemble, 
provided the time of observation is long enough.
The second route is to characterize the properties of 
{\em individual} systems by means of the periodic orbit theory
\cite{Gutzwiller}. This is possible for a system with chaotic 
dynamics governed by a Hamiltonian that is simple enough, so that 
the parameters of the periodic orbits needed for semiclassical spectral
analysis can be calculated explicitly.
Averaging in this case is usually done over an energy interval 
which consists of many energy levels \cite{Berry85}.
This approach is very powerful in describing the short time 
behavior of the system, but is faced with significant problems
when applied to times of order or bigger than the 
Heisenberg time $\tau_H=h/\Delta$ or to energies much smaller 
than the mean level spacing $\Delta$ \cite{Wilkinson}.  
Despite the obvious differences between the two approaches 
it is believed that to a large extent both describe 
the same physics. In this letter we are applying results of 
the first approach in order to extend the periodic orbit 
theory to times close to $\tau_H$. 

\par 

The relation between ensembles of metallic grains and ensembles
of random matrices (RM) \cite{Mehta91} is now well understood.
The supersymmetric non-linear $\sigma$ model \cite{Efetov83}, 
actually, provides a microscopic justification for the use 
of RM theory in order to describe the universal features of 
these systems. This formalism offers a routine way of 
calculation of a variety of universal correlation functions
for all Dyson pure symmetry RM ensembles and for crossovers
between them \cite{Les-Houches}. In view of the growing 
interest in applying the super-symmetry approach to the 
investigation of deviations from universality
\cite{Kravtsov,Andreev,Altland}, it becomes  
important to understand the connection between the two 
approaches described above. 

The object that we analyse  
is the dimensionless two--point level density correlator,
\begin{equation}
R(s) =\Delta^2\langle \rho(E) \rho(E+s\Delta) \rangle
 - 1. \label{correlator}
\end{equation}  
Here $\rho(E)$ is the density of states at energy $E$,
$\Delta$ is the mean level spacing, and $\langle \cdots \rangle$
represents ensemble averaging in the case of disordered systems,
or averaging over some interval of energy $E$
if an individual chaotic system is considered. The universal
form of $R(s)$ is especially simple in the unitary case.
It is the sum of smooth  and oscillating parts 
\cite{Mehta91}: $R(s)=\delta(s)-[1-\cos(2\pi s)]/(2\pi^2 s^2)$. The
conventional perturbation theory for disordered metals
\cite{Abrikosov} can provide only the smooth part of $R(s)$ 
\cite{Altshuler85}.
The $s\gg 1$ asymptotics of $R(s)$ in which the oscillatory
term, non-analytic in $1/s$, is retained was recently evaluated 
in Ref. \cite{Andreev}. This result (for $s\neq 0$) can be 
still presented as a sum, 
\begin{equation}
R(s)=R_P(s)+R_{osc}(s), 
\label{rpertosc}
\end{equation}
of a perturbative term 
$R_P(s)$ and oscillatory one $R_{osc}(s)$. We 
rewrite the expression for $R_P(s)$ (see Ref.
\cite{Altshuler85}) as 
\begin{equation}
R_P(s)= -\frac{1}{2\alpha \pi^2} \frac{\partial^2}{\partial s^2}
\ln [{\cal D}(s)], \label{perturbative}
\end{equation} 
where $\alpha=2$ for the unitary ensemble and $\alpha=1$ for
T-invariant ensembles. ${\cal D}(s)$ is the spectral determinant 
of a classical operator, namely the diffusion operator:
\begin{equation}
{\cal D}(s)= \prod_{\mu} A(\epsilon_{\mu}) \left( s^2+\epsilon_{\mu}^2
\right)^{-1}. \label{Anton-spectral}
\end{equation} 
Here $\epsilon_{\mu}$ are eigenvalues (in units of $\Delta$)
of the diffusion equation in the grain, and $A(\epsilon_{\mu})$
is a regularization factor which equals $\epsilon_{\mu}^2$ for
$\epsilon_{\mu}\neq 0$ and unity otherwise~\cite{note}. 
Surprisingly the
oscillatory term $R_{osc}(s)$, which cannot be obtained
by a perturbative calculation, is also  governed by the same 
classical spectral determinant ${\cal D}(s)$. E. g., in the unitary
case it has the form
\begin{equation}
R_{osc}(s)= \frac{\cos(2\pi s)}{2\pi^2} {\cal D}(s). 
\label{unitary} 
\end{equation}
Since ${\cal D}(s)$ is purely classical, it is plausible that for any 
chaotic system there exists a classical operator whose 
spectral determinant can be identified with ${\cal D}(s)$. 
In what follows we shall identify this operator for general 
chaotic systems by a semiclassical analysis of relation
(\ref{perturbative}). For the sake of simplicity we shall 
consider a two dimensional system which belongs 
to the unitary ensemble.

The semiclassical analysis begins with  
Gutzwiller's trace formula \cite{Gutzwiller}, which expresses 
the density of states $\rho(E)$ as a sum over the 
classical periodic orbits

\begin{equation}
\rho(E)= \frac{1}{\Delta}+\Re \frac{1}{\pi\hbar} 
\sum_{p} T_p \sum_r \frac{e^{\frac{i}{\hbar}S_{p}(E)r 
-i\nu_pr}}{ |\det (M_p^r-I) |^{1/2}}, \label{Gutzwiller}
\end{equation} 
where $p$ labels a primitive orbit that is characterized by a
period $T_p$, action $S_p(E)$, and Maslov phase 
$\nu_p$; $r$ stands for the number of the repetitions of this 
orbit. $M_p$ is the monodromy matrix associated with the 
linearized dynamics on the Poincar\`{e} section perpendicular 
to the orbit. From here on, energy and time will be measured 
in units of $\Delta$ ($\epsilon =E/\Delta$), 
and $\hbar/\Delta$ respectively.
One can substitute (\ref{Gutzwiller}) into
(\ref{correlator}) and represent $R(s)$ in the form of a 
double sum over the periodic orbits. $R_P(s)$ is given by 
the diagonal part of this sum. Expanding $S_p(\epsilon+s)$
up to the linear order in $s$: $S_p(\epsilon+s)\simeq  S_p(\epsilon)
+T_p s$, we obtain
\begin{equation}
R_P(s)= \Re \frac{1}{2\pi^2} \sum_p T_p^2 \sum_{r=1}
^{\infty} \frac{ e^{isT_p r}}{
|\det (M_p^r-I)|}. \label{diagonal}
\end{equation} 
The traditional way to deal with the above sum is to 
approximate it by an integral:
\begin{equation}
\sum_p \frac{f(T_p)}{|\det (M_p-I)|} \to \int \frac{dt}{t}f(t)
\end{equation} 
for any sufficiently smooth function $f(t)$.
This approximation, known as the Hannay and 
Ozorio de Almeida (H\&OA) sum rule, holds in the limit 
$t\to \infty$ where long periodic orbits 
which explore the whole energy shell uniformly
are considered. In employing it for the calculation
of $R_P(s)$, the time $t$ should be restricted to the regime
where it is much larger than the shortest periodic orbits 
but still smaller than the Heisenberg time $\tau_H$. The result
associated with it is therefore the universal one $R_P(s)
=-1/2\pi^2s^2$ which holds as long as $s \gg 1$ \cite{Berry85}.
Below we present a more careful treatment of the sum
(\ref{diagonal}) that keeps the non-universal part of $R_P(s)$.

\par

Let $\Lambda_p$ be the eigenvalue ($|\Lambda_p|>1$)
of the monodromy matrix $M_p$. The area preserving property 
of the latter implies that the second eigenvalue of $M_p$ is 
$1/\Lambda_p$. Hence,

\begin{equation}
|\det (M_p^r-I)|^{-1}= |\Lambda_p|^{-r}\sum_{k=0}^{\infty} 
(k+1) \Lambda_p^{-rk},
\end{equation}
and we can rewrite (\ref{diagonal}) in the
form of a triple sum
\begin{equation}
R_P(s)= \frac{-1}{2\pi^2} \frac{\partial^2}
{\partial s^2} \Re \sum_{p,k}(k+1)\sum_{r=1}^{\infty}
\frac{1}{r^2}t_{pk}^r,
\label{triplesum}
\end{equation}
where 
\begin{equation}
t_{pk}=|\Lambda_p|^{-1}\Lambda_p^{-k}e^{isT_p}.
\end{equation}

\noindent Using the relation (\ref{perturbative}) we can determine 
the spectral determinant ${\cal D}(s)$ up to a normalization constant: 
\begin{equation}
\label{D-zetas}
{\cal D}(s)= |{\cal N}\tilde{Z}(is)|^2. \label{spectral-zeta}
\end{equation}
Upon evaluation of the sum over the repetitions in Eq.~(\ref{triplesum}), 
the expression for $\tilde{Z}(is)$ takes the form 
\begin{equation}
1/\tilde{Z}(is)= \prod_p\prod_{k=0}^{\infty}
\exp \left[(k+1)\phi(t_{pk})\right],
\end{equation}
where $\phi (x)=\int_{0}^{x}t^{-1}\ln(1-t)dt$.
Notice that the normalization constant ${\cal N}$ plays no
role in the perturbative part of the two--point correlator.
We therefore postpone its determination.

\par

Suppose now that all the periodic orbits are
very unstable, namely $|\Lambda_p|\gg 1$ for all $p$-s.
In this case $t_{pk} \to 0$, $\phi(t_{pk})\to -t_{pk}$ 
and $\tilde{Z}(z)$ reduces to the dynamical zeta function \cite{Ruelle},
\begin{equation}
1/Z(z)= \prod_p\prod_{k=0}^{\infty} \left( 1-
\frac{e^{zT_p}}{|\Lambda_p| \Lambda_p^k}\right)^{k+1}. 
\label{Ruelle}
\end{equation}
This function is the spectral determinant associated with
the Perron-Frobenius (PF) operator ${\cal L}^t$ (also known 
as Ruelle-Araki or the transfer operator) \cite{Cvitanovic}. 
${\cal L}^t$ is the classical evolution operator which 
propagates phase space density for a time $t>0$. Its 
kernel is therefore given by 
\begin{eqnarray}
{\cal L}^t( \vec{y},\vec{x})= \delta [\vec{y}-\vec{u}
(\vec{x};t)], 
\end{eqnarray}
where $\vec{y}$ and $\vec{x}$ are phase space vectors 
representing coordinates and momenta, 
and  $\vec{u}(\vec{x};t)$ is the point in phase 
space to which a particle that starts its motion
at $\vec{x}$ arrives after time $t$. The eigenvalues of 
the PF operator are of the form $e^{-\gamma_{\mu}t}$. They are
associated with the decaying modes of a disturbance in the 
density of classical particles exhibiting chaotic dynamics, 
analogous to the diffusion modes of disordered system. Yet, 
the difference is that, unlike in the latter case, here  
$\gamma_{\mu}$-s can appear also in complex conjugate pairs
$\gamma\!=\!\gamma^{\prime}\pm i\gamma^{\prime\prime}$ where
$\gamma^{\prime}\geq 0$. The leading eigenvalue 
of the PF operator, $\gamma_{0}\!=\!0$, corresponds to the 
conservation of the number of particles. The dynamical zeta 
function (\ref{Ruelle}) is the spectral determinant associated with the 
eigenvalues $\gamma_{\mu}$. Up to a normalization constant 
it is given by the product
\begin{equation}
1/Z (z) = \prod_{\mu} B_{\mu} (z-\gamma_{\mu}),
\label{Ruelle-prod}
\end{equation}
where $B_{\mu}$ are regularization factors 
introduced to make the product converge.

Unlike the periodic orbit theory in quantum mechanics which
gives only the leading asymptotics in the limit $\hbar \to 0$, 
the periodic orbit expansion (\ref{Ruelle}) of (\ref{Ruelle-prod}) 
is exact. It is however proper to comment that, in its present 
form, $Z(z)$ cannot be used to determine the eigenvalues 
$\gamma_{\mu}$. For this purpose a resummed formula is required. 
It can be obtained by expanding the infinite product over 
the periodic orbits and ordering the various terms in a way 
that leads to  maximal cancelation among them. This so called 
cycle expansion \cite{Artuso90} exploits the property that 
the dynamics of chaotic systems in phase space 
is coded by a skeleton of few periodic orbits. In particular, 
the long periodic orbits may be approximated by 
linear combinations of few short ones.

\par

>From (\ref{Ruelle-prod}) and (\ref{perturbative}) it follows that
\begin{equation}
R_P(s)= \Re \frac{1}{2\pi^2} \sum_{\mu} 
\frac{1}{(-is +\gamma_{\mu})^2} \label{rpert}
\end{equation}
in complete analogy with the result of Ref.~\cite{Altshuler85} 
for diffusive systems.  
The universal part of $R_P(s)$, that was obtained
using H\&OA sum rule, thus corresponds to the 
first term in the sum (\ref{rpert}) ($\gamma_0=0$). 
The rest of the sum is apparently system-specific.

\par

We turn now to the determination of the normalization constant 
${\cal N}$ introduced in (\ref{spectral-zeta}). We shall 
assume that the leading eigenvalue $\gamma_0$ is 
of unit multiplicity (this is the case when the 
system is ergodic).  Comparison of Eqs.~(\ref{spectral-zeta}) 
and (\ref{Anton-spectral}) gives the normalization factor:
\begin{equation}
{\cal N}^{-1} = \lim_{z\to 0} z Z(z).
\end{equation}

\par

It is customary to express the semiclassical density of states  
as the logarithmic derivative of the Selberg zeta function.
The latter is defined as the spectral determinant associated 
with the semiclassical energy spectrum of the system under 
consideration:
\begin{equation}
\zeta_s(\epsilon) = \prod_j b_j
(\epsilon-\epsilon_j)= \prod_p\prod_{k=0}^{\infty} 
\left( 1- \frac{e^{iS_p(\epsilon)-i\nu_p}}{|\Lambda_p|^{1/2} 
\Lambda_p^k}\right), 
\label{rhozeta}
\end{equation} 
where $b_j$ are regularization factors, and
$\epsilon_j$ are the semiclassical energy levels of the system.
The second equality above holds for two dimensional systems. 
One can show that the spectral determinant $\tilde{Z}(is)$, 
satisfies the relation
\begin{equation}
\label{speczetas}
\tilde{Z}(is)=\exp \{ \langle \ln [\zeta_s(\epsilon+s)]
 \ln[\zeta^*_s(\epsilon)]\rangle_{\rm d} \}. 
\end{equation}
where $\langle \cdots \rangle_{\rm d}$ represents an
averaging which retains only the diagonal elements 
in the double sum.
Since $\Delta \rho(\epsilon)=1-(\partial/\pi\partial \epsilon)\Im 
\ln \zeta_s(\epsilon+i0)$,  the two-point correlator can be 
written as
\begin{equation}
  \label{Rexact}
  R(s)=\frac{-\partial^2}{\pi^2\partial s^2}
  \left\langle
  \Im \ln \zeta_s\left(\epsilon+s\right)
  \Im \ln \zeta_s\left(\epsilon \right)
\right\rangle.
\end{equation}
The diagonal approximation gives the perturbative term
\begin{equation}
  \label{Rdiag}
  R_P(s)=\frac{-\partial^2}{\pi^2\partial s^2}
  \left\langle
  \Im \ln \zeta_s\left(\epsilon+s\right)
  \Im \ln \zeta_s\left(\epsilon\right)
\right\rangle_{\rm d}.
\end{equation}
The difference between Eqs.~(\ref{Rexact}) and (\ref{Rdiag}) 
can be also expressed through the diagonal average.
Using Eqs.~(\ref{speczetas}) and (\ref{D-zetas}) it is easy to see 
that $R_{osc}(s)$ is given by Eq.~(\ref{unitary}) with 
\begin{equation}
{\cal D}(s)= {\cal N}^2 \exp \left\{ 2 \Re \left\langle
\ln \zeta_s\left(\epsilon+s\right)
\ln \zeta_s^*\left(\epsilon \right)
\right\rangle_{\rm d} \right\}.
\end{equation}

It is convenient to present the result in terms of the 
Fourier transform of the two--point level density correlator,
$S(\tau)=\int ds e^{is\tau}R(s)$, known as the structure or
the form factor. RM theory predicts that for the unitary 
ensemble $S(\tau)=min( |\tau|/2\pi, 1)$ (see the light line 
in Fig. 1). In the general case
\begin{equation}
S(\tau)= S_P(\tau)+ \frac{1}{2}\left[ S_{osc}(\tau+2\pi)+
S_{osc}(\tau-2\pi)\right],
\end{equation}
where $S_P$ and $S_{osc}$ are respectively associated with the 
perturbative (\ref{perturbative}) and the non-perturbative
(\ref{unitary}) parts of the two--point correlator.
Assuming that the multiplicity of all the eigenvalues 
$\gamma_{\mu}$ is one, 
\begin{equation}
S_P(\tau)= \frac{|\tau|}{2\pi} \sum_{\mu} e^{-\gamma_{\mu} |\tau|}.
\label{Perturbative-form}
\end{equation}
Again the universal part of $S_P(\tau)$ associated with H\&OA
sum rule comes from the leading eigenvalue $\gamma_0=0$. 
The higher eigenvalues will contribute corrections which 
are in general oscillatory and decrease exponentially. 
For instance, the complex pair $\gamma_1^{\prime}
\pm i\gamma_1^{\prime \prime}$ will contribute
the term $|\tau| e^{-\gamma_1^{\prime}|\tau|} 
\cos (\gamma_1^{\prime \prime}\tau)/\pi$.  
The oscillatory part of the structure factor
can be written as,
\begin{equation}
S_{osc}(\tau)= -\frac{|\tau |}{2\pi} -\sum_{\mu \neq 0}
\frac{{\cal D}_{\mu}(i\gamma_{\mu})}{2\pi\gamma_{\mu}} e^{-\gamma_{\mu} 
|\tau|},
\end{equation} 
where ${\cal D}_{\mu}(s)$ is given by
\begin{equation}
{\cal D}_{\mu}(s)= \left(1+\frac{s^2}{\gamma_{\mu}^2}\right) 
s^2 {\cal D}(s).
\end{equation}
For example, in the case of quasi one dimensional diffusive system, 
where the eigenvalues are of the form $\gamma_{n}\!=\!Dn^2$ one can
show that ${\cal D}_{n}(iDn^2)\!=\! -4n(-1)^n/\sinh (\pi n)$, while for
equally spaced eigenvalues $\gamma_{n}\!=\!vn$ it is  ${\cal D}_{n}(ivn)
=2\pi n/\sinh(\pi n)$. In general it is expected that the
contribution will come only from the lowest eigenvalues of the
PF operator. 

In what follows it will be assumed that the non-universal
behaviour is dominated by one eigenvalue (or possibly a 
conjugate pair) $\gamma_1$, i.e. $\gamma_{\mu}^{\prime}\!\gg\!
\gamma_1^{\prime}$ for all $\mu\!>\!1$.
  
\begin{figure}
  \begin{center}
 \leavevmode
    \epsfxsize=6cm 
\epsfbox{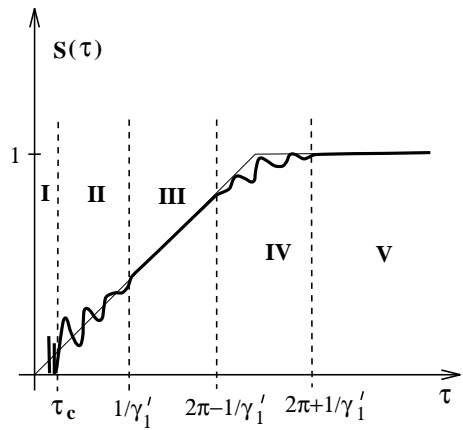}
 \end{center}
  \caption{A schematic draw of the structure factor
     of chaotic system belonging to the unitary ensemble.
     The light line represents the universal RM
     theory result. For the sake of clarity, the non-universal 
     features have been exaggerated.}
  \label{fig:1}
\end{figure}

In characterizing $S(\tau)$, five domains of the parameter $\tau$,
drawn schematically in Fig 1, are identified:
(I) $\tau \sim \tau_c$ where $\tau_c$ is of order of
the period of the shortest periodic orbit. Here $S(\tau)$ is 
composed of several $\delta$-function peaks located at the periods 
of the short orbits and weighted according to their instability. 
(II) $\tau_c < \tau < 1/\gamma_1^{\prime}$. Deviations from
universality associated with (\ref{Perturbative-form}) may be 
noticeable also in this interval. Their period of oscillation $1/
\gamma_1^{\prime \prime}$ is of the order of $\tau_{c}$.
(III)  $ 1/\gamma_1^{\prime} \!<\! \tau \!
< \! 2\pi-1/\gamma_1^{\prime}$,  the universal perturbative
regime where $S(\tau) = \tau/2\pi$. This is the domain where
H\&OA sum rule holds. 
(IV) $ 2\pi -1/\gamma_1^{\prime}\!<\! 
\tau \!< \! 2\pi+1/\gamma_1^{\prime}$,  the vicinity of the 
Heisenberg time $\tau=2\pi$. The non-universal features here
are in the form of exponentially decreasing oscillations
very similar to those existing in (II). Yet their amplitude 
and phase may be different. In general, the RM singularity 
at the Heisenberg time (the light line in Fig. 1) will be 
smeared out by them. 
(V) $\tau > 2\pi+1/\gamma_1^{\prime}$. Here again the 
universal result $S(\tau)=1$ holds.

These results can be generalized straightforwardly to 
orthogonal and symplectic chaotic systems: instead of 
Eq.~(\ref{unitary}) one should use Eq.~(5) of 
Ref.~\cite{Andreev} with $P(s)=s^2{\cal D}(s)$.

The behavior of the structure 
factor in the vicinity of the Heisenberg
time is a manifestation of a striking property of the
periodic orbit sum (\ref{Gutzwiller}). Namely, that the tail of
the Gutzwiller's series (the long periodic orbit) encodes 
its head (short periodic orbit). As a result, $S(\tau)$ in the 
vicinity of the Heisenberg time is determined by the same 
short periodic orbits as at small $\tau$.
The argument \cite{Berry85}
is that the long periodic orbits determine the position
of the energy levels. Therefore through the long range
correlation of these levels they encode the information about
the short periodic orbits. In fact, Berry and Keating 
resummation method \cite{Berry92} 
of the periodic orbit sum associated with the quantum 
spectral determinant of chaotic systems is based on the 
bootstrapping of long periodic orbits with periods near
the Heisenberg time $\tau\sim2\pi$ to the short ones near 
$\tau_{c}$. The behavior of $S(\tau)$ near the Heisenberg 
time reflect this sort of symmetry in the sense that it is 
determined by the short time dynamics of the classical system.

To summarize, we identified the diffusion operator in disordered
grains with the Perron-Frobenius operator in the
general case. This relates the spectral determinant associated
with the diffusion equation in the grain to the 
dynamical zeta function which can be expressed in terms of the 
classical periodic orbits. We used these relations 
to extend the theory of the structure factor
of disordered grains to general chaotic systems.
It would be interesting and important to derive these relations
for generic chaotic systems. In this respect the recently proposed 
$\sigma$-model-like approach for ballistic systems \cite{Muzykantsky95} 
looks promising.
 
\par    
{\it Acknowledgements:} 
It is our great pleasure to thank P.~Cvitanovi\'c, 
B.~Eckhardt, D.~E.~Khmelnitskii, B.~A.~Muzykantsky, 
D.~Ruelle and B.~D.~Simons, for 
very informative discussions. This work was supported in part 
by the US-Israel Binational Science Foundation (BSF) and 
by the NSF Grant No. DMR 92-04480.
  

\begin{thebibliography}{99}   
\bibitem{Gutzwiller} M. C. Gutzwiller, J. Math. Phys. {\bf 8}, 
1979 (1967); {\bf 10}, 1004 (1969); {\bf 11}, 1791 (1970); 
{\bf 12}, 343 (1971). 
\bibitem{Berry85} M. V. Berry, Proc. Roy. 
Soc. London {\bf A 400}, 229 (1985).
\bibitem{Wilkinson} M. Wilkinson, J. Phys. A {\bf 21}, 
1173 (1988). 
\bibitem{Mehta91} M. L. Mehta, Random Matrices, 
(Academic Press, New York, 1991)
\bibitem{Efetov83} K. B. Efetov Adv. Phys. {\bf 32}, 53 (1983).
\bibitem{Les-Houches} B. L. Altshuler and B. D. Simons, in
Proceedings of Les-Houches Summer School, session LXI, 
1994, Eds.  E. Akkermans, G. Montambaux, J-L. Pichard, and 
J. Zinn-Justin, to be published. 
\bibitem{Kravtsov} V. E. Kravtsov and A. D. Mirlin, Sov. Phys.
JETP Lett {\bf 60} 656 (1994). [Pis'ma ZhETF, {\bf 60},645 (1994)].
\bibitem{Andreev} A. V. Andreev and B. L. Altshuler, 
Phys. Rev. Lett. {\bf 75}, 902 (1995).
\bibitem{Altland} A. Altland and D. Fuchs,  Phys. Rev. Lett.
{\bf 74}, 4269 (1995).
\bibitem{Abrikosov} A. A. Abrikosov, L. P. Gor'kov, and 
I. E. Dzyaloshinski, Methods of Quantum Field Theory 
 in Statistical Physics (Dover, New York, 1975).
\bibitem{Altshuler85}B. L. Altshuler and B. I. Shklovskii, JETP
{\bf 64}, 127 (1986).
\bibitem{note}
Note that ${\cal D}(s)$ differs from $P(s)$ of 
Ref.~\cite{Andreev} by 
inclusion of the zero mode: ${\cal D}(s)=P(s)/s^2$. 
\bibitem{Hannay} J. H. Hannay and A. M. Ozorio 
de Almeida, J. Phys. A {\bf 17}, 3429 (1984).
\bibitem{Ruelle} D. Ruelle, Statistical Mechanics, 
Thermodynamic Formalism (Reading, MA: Addison-Wesley, 1978).
\bibitem{Cvitanovic} P. Cvitanovi\'{c} and B. 
Eckhardt, J. Phys. A: Math Gen {\bf 24}, L237 (1991).
\bibitem{Artuso90} R. Artuso, E. Aurell and P. Cvitanovi\'{c},
Nonlinearity {\bf 3}, 325 (1990); Nonlinearity {\bf 3}, 361 (1990).
\bibitem{Berry92} M. V. Berry and J. P. Keating, 
Proc. R. Soc. Lond. A {\bf 437} 151 (1992).
\bibitem{Muzykantsky95}B. A. Muzykantsky, D. E. Khmel'nitskii,  
JETP Lett. (1995)
\end{thebibliography}
\end{document}